\documentstyle[prl,aps,preprint,epsf]{revtex}

\begin{document}
\newcommand{\be}{\begin{equation}}
\newcommand{\ee}{\end{equation}}
\newcommand{\bq}{\begin{eqnarray}}
\newcommand{\eq}{\end{eqnarray}}
\newcommand{\Sc}{Schr\"odinger\,\,}
\newcommand{\Sp}{\,\,\,\,\,\,\,\,\,\,\,\,\,}
\newcommand{\no}{\nonumber\\}
\newcommand{\tr}{\text{tr}}
\newcommand{\p}{\partial}
\newcommand{\la}{\lambda}
\newcommand{\G}{{\cal G}}
\newcommand{\D}{{\cal D}}
\newcommand{\W}{{\bf W}}
\newcommand{\de}{\delta}
\newcommand{\al}{\alpha}
\newcommand{\bi}{\beta}
\newcommand{\ga}{\gamma}
\newcommand{\ep}{\epsilon}
\newcommand{\vep}{\varepsilon}
\newcommand{\th}{\theta}
\newcommand{\om}{\omega}
\newcommand{\J}{{\cal J}}
\newcommand{\pr}{\prime}
\newcommand{\ka}{\kappa}
\newcommand{\TH}{\mbox{\boldmath${\theta}$}}
\newcommand{\DE}{\mbox{\boldmath${\delta}$}}

\setcounter{page}{0}
\def\footnoterule{\kern-3pt \hrule width\hsize \kern3pt}
\tighten
\title{
The Coulomb Branch of Yang-Mills Theory from the Schr\"odinger Representation
}
\author{Jiannis Pachos
\footnote{Email address: {\tt pachos@ctp.mit.edu}}
}
\address{Center for Theoretical Physics \\
Massachusetts Institute of Technology \\
Cambridge, Massachusetts 02139 \\
{~}}

\date{MIT-CTP-2710, January 1998}
\maketitle

\thispagestyle{empty}

\begin{abstract}

The Coulomb branch of the potential between two static colored sources is calculated for the Yang-Mills theory using the electric Schr\"odinger representation.

\end{abstract}

%\vspace*{\fill}

%\newpage

\section{Introduction}

We shall study the $(3+1)$ dimensional Yang-Mills theory with static sources in the Electric Schr\"odinger representation. The static sources are placed in order to visualize the behavior of the non-Abelian field \cite{Rossi}. We aim to extract in an easy way the Coulomb branch of the potential of the non-Abelian $SU(N)$ theory. The method adopted here is a generalization of the one used for the equivalent $(1+1)$-dimensional case in \cite{AJ} and \cite{Jiannis}. The Coulomb branch can be derived also in the A-representation with the use of the Wilson loop, but the steps there are approximate and require assumptions. The method presented here is straightforward. It focuses on the kinetic part of the Hamiltonian, which will give, as its expectation value, the Coulomb potential between the static sources. Further study in this direction is needed to obtain the linear confining potential.

\section{Wave Functional and its Transformation} 

In \cite{Nair} the functional 
\be
\Psi[E]=\int \D u e^{-{1 \over c} \tr \int E^i \p_iu u ^{-1}}\Phi[E]
\label{nair1}
\ee
is presented as a solution of the free Gauss' law, where $\Phi[E]$ is a gauge invariant functional of the electric field, $E$. In the case we have two static sources (source - anti-source) placed at the points $x_0$ and $x_1$ the Gauss' law becomes
\be
G_a(x)\Psi[E]=\left(\p _iE_a^i(x)-if_{abc}E_b^i(x) {\de  \over \de E_c^i(x)}\right)\Psi[E]=\Psi[E]T_a \de(x-x_0)-T_a\Psi[E] \de(x-x_1)
\label{ggg}
\ee
and the wave functional satisfying it is
\be
\Psi[E]=\int \D u e^{-{1 \over c} \tr \int E^i \p_iu u ^{-1}} u(x_1)u^{-1}(x_0) \,\,\Phi[E]
\label{nair2}
\ee
There could be a constant matrix between $u(x_1)$ and $u^{-1}(x_0)$ and still satisfying (\ref{ggg}), as $u$ transforms like $u\rightarrow Uu$ under a $U \epsilon \, SU(N)$ transformation. Under the gauge transformation, $E \rightarrow E^U$, this wave functional transforms as
\be
\Psi[E^U]=e^{-{1 \over c}\tr \int E^i \p _iU^{-1} U} U(x_1)\Psi[E] U^{-1}(x_0)
\ee
This is the generalization to the case with sources of the gauge transformation with the wave functional in the E-representation \cite{Jackiw} \cite{Freedman}.

\section{Properties of the Wave Functional}

As the gauge transform of $E^i$ is $\left. E^i\right. ^U=UE^iU^{-1}$, we can decompose the electric field as $E^i=gK^ig^{-1}$. $g \in SU(N)$ and transforms as $g \rightarrow Ug$, while $K^i$ are matrices belonging in ${\it su}(N)$ and they do not transform under $U$. The algebraic components of $K^i$ are given by $K^i=\sum_{a=1}^{N^2-1}k^i_a \,T^a$. The fixed direction of $K^i$ in the $SU(N)$ space makes the theory to reduce to an Abelian theory with respect to $K^i$. A specific direction of it can be chosen by taking the average vector $\sum_i K^i$ to be along the Cartan sub-algebra.

With this decomposition the functional (\ref{nair2}) becomes
\be
\Psi[E]=e^{- {1 \over c} \tr \int E^i \p_ig g^{-1}} g(x_1) \left[ \int \D u e^{- {1 \over c} \tr \int K^i \p_iu u^{-1}} u(x_1) u^{-1} (x_0) \right] g^{-1} (x_0) \,\, \Phi[E]\,\, .
\label{inter}
\ee
We shall use the decomposition of the $SU(N)$ group presented in \cite{Jiannis} of the form $u=h(\phi)\tilde u(\th) h(\bar \phi)$, where the elements, $h$, belong in the Cartan subgroup. The parameterization in $N-1$ angles $\phi^k$, $N-1$ angles $\bar \phi^k$ and the rest $(N-1)^2$ angles $\th$, is such that
\be
{\cal J}_k u\equiv -i{\p \over \p \phi^k}u=-T^k u \,\,\,\,\, , \,\,\,\,\,\,\,\,\,\,\,\,\,\,\,\,\,\, {\cal J}^R_k u \equiv -i{\p \over \p \bar \phi^k}u = u T^k \,\, ,
\ee
where $T^k$ are the $N-1$  Cartan generators. They can be chosen to be diagonal, with $f_k(s)$ their $s$ element. With these ``diagonalization'' conditions the group integration measure $\D u$ becomes $[\D\phi][\D\bar \phi][\D\th]J(\th)$, where $J(\th)$ is the Jacobian of the reparameterization and depends only on the angles, $\th$. This allows to calculate relatively easy the integrations in relation (\ref{inter}), obtaining finally
\bq
\Psi[E]=e^{- {1 \over c} \tr \int E^i \p_ig g^{-1}} &&\sum_\rho  g(x_1) P^\rho g^{-1} (x_0) 
\no \no
&&
\times \prod_{k=1}^{N-1} \prod_x \de \Big( \p_i k^i_k-f_k(\rho)\de (x-x_0) +f_k(\rho) \de (x-x_1) \Big) \,\, \Phi[E] \,\, .
\label{abel}
\eq
The projection matrices, $P^\rho$, are diagonal with $1$ in the $\rho$ position and zeros everywhere else. From (\ref{abel}) the Abelian character of the theory with respect to $K^i$ is apparent. The complete wave functional is a superposition of $N+1$ functionals like $\Psi[E]$ corresponding to the $N+1$ directions the Cartan sub-algebra could take in the $SU(N)$ algebra.

The delta functions in the above expression force $k_k^i$ to be the $x_i$ derivative of the Green's function $G(x)$, which is a solution of the following equation
\be
\p_i \p^i G(x) =f_k(\rho)\de(x-x_0)-f_k(\rho)\de(x-x_1)
\ee
For $(d+1)$ dimensions $k^i$ is given by
\be
k^i_k (x)= \Sp \left\{
\begin{array}{ll}
\Sp f_k(\rho)[\th(x-x_0)-\th(x-x_1)]& \Sp \mbox{for $d=1$,} \\ 
\Sp {f_k(\rho) \over 2 \pi} \p_i \left(\log {|x-x_0| \over |x-x_1|} \right) & \Sp \mbox{for $d=2$,} \\ 
\Sp-{f_k(\rho) \over (d-2)\omega_d} \p_i \Big({1 \over |x-x_0|^{d-2}}-{1 \over |x-x_1|^{d-2}} \Big) & \Sp \mbox{for $d>2$,}
\end{array}
\right.
\label{green}
\ee
where $\omega_d=2\pi^{d/2}/\Gamma (d/2)$. For $d=1$ the results in \cite{Jiannis} are reproduced.
In the following we shall substitute the previous delta functions in $\Psi[E]$ with ones which impose these equalities. 

\section{Coulomb Potential}

For the $(3+1)$-dimensional Yang-Mills theory the expectation value of the Hamiltonian \cite{Paul} with respect to the $\Psi[E]$ state is given by
\be
V=V_{kin}+V_{pot}={1 \over 2} \int dx \int \D E \Psi^{+}[E]\Big(g^2E^i_a(x) E^i_a(x) + {1 \over g^2}B^i_a(x)B^i_a(x)\Big) \Psi[E]
\ee
where $B$ is the non-Abelian magnetic field. $B$, as a function of the vector potential $A^i_a=i{\de \over \de E^i_a}$, includes up to fourth order functional differentiations. Here, we shall restrict ourselves on the kinetic term, $\sum_{a=1}^{N^2-1} k^i_a k^i_a=\sum_{k=1}^{N-1} k^i_k k^i_k + \sum_{\la=N}^{N^2-1} k^i_{\la} k^i_{\la}$. The first part is easy to calculate due to the delta functions in $\Psi[E]$, while, $\sum_{\la=N}^{N^2-1} k^i_{\la} k^i_{\la}$, will give an overall additive constant. Finally, the following kinetic potential is obtained
\be
V_{kin}= V^0_{se}+V^1_{se} - g^2 C_2^{(N-1)} {1 \over 4 \pi} {1 \over |x_1 - x_0|}
\ee
where $V^0_{se}$ and $V^1_{se}$ are the Coulomb self-energies of the sources at the points $x_0$ and $x_1$, and $C_2^{(N-1)}$ is the quadratic Casimir operator restricted {\it only} on the Cartan components, i.e. $\sum_{k=1}^{N-1}T^kT^k=C_2^{(N-1)} {\bf 1}= C_2/(N+1) {\bf 1}$. A sum over the $N+1$ directions the Cartan subalgebra could take will result to the complete Coulomb potential
\be
V^{01}_{kin}=-g^2C_2 {1 \over 4 \pi} {1 \over |x_1 - x_0|} \,\, .
\ee
In $(1+1)$ dimensions the theta functions in (\ref{green}) produce a confining potential.

\section{Conclusions}

With straightforward steps we have calculated the Coulomb branch of the potential of two static non-Abelian sources. It is extracted from the kinetic term of the Hamiltonian which is proportional to $g^2$. The more interesting magnetic part is proportional to $1/g^2$ and is expected that after a proper regularization (see e.g. \cite{Paul} \cite{MSP} \cite{PJ}) its calculation will produce the linear potential between the static sources.

\end{document}